\documentclass{optica-article}

\journal{opticajournal}

\articletype{Research Article}

\usepackage{lineno}

\begin{document}

\title{Continuous optical generation of microwave signals for fountain clocks}

\author{Burghard~Lipphardt, Patrick~Walkemeyer, Michael~Kazda, Johannes~Rahm and~Stefan~Weyers\authormark{*}}

\address{Physikalisch-Technische Bundesanstalt (PTB), Bundesallee 100, 38116 Braunschweig, Germany}

\email{\authormark{*}stefan.weyers@ptb.de}

\begin{abstract*} 
For the optical generation of ultrastable microwave signals for fountain clocks we developed a setup, which is based on a cavity stabilized laser and a commercial frequency comb. The robust system, in operation since 2020, is locked to a 100\,MHz output frequency of a hydrogen maser and provides an ultrastable 9.6\,GHz signal for the interrogation of atoms in two caesium fountain clocks, acting as primary frequency standards. Measurements reveal that the system provides a phase noise level which enables quantum projection noise limited fountain frequency instabilities at the low $10^{-14} (\tau /\mathrm{s})^{-1/2}$ level. At the same time it offers largely maintenance-free operation.

\end{abstract*}

\section{Introduction}
\label{sec_intro}

The performance of atomic clocks is characterized by their systematic uncertainty and frequency instability. Depending on the actual measurement time the latter determines the statistical measurement uncertainty. Today the most accurate realization of the SI second is obtained from caesium fountain clocks \cite{Wynands2005}. In the best case systematic and statistical measurement uncertainties at the low $10^{-16}$ level have been reached in measurement campaigns lasting a number of days \cite{CircT}. Here advanced methods of the microwave generation, making use of cryogenic oscillators or ultrastable lasers, are advantageous \cite{Santarelli1999,Lipphardt2009,Millo2009a,Takamizawa2014}, which allow to overcome the otherwise limiting phase noise of microwave syntheses based on commercially available quartz oscillators. 
Fountain clock applications which benefit from the lowest overall uncertainties are calibrations of International Atomic Time (TAI) \cite{CircT}, absolute frequency measurements of optical clock transitions \cite{Schwarz2020,Lange2021,Nemitz2021}, steering of local time scales \cite{Bauch2012,Rovera2016} and fundamental physics, like the search for changes in fundamental constants \cite{Lange2021,Schwarz2020,Godun2014}, violations of local position or Lorentz invariance (LPI and LLI) \cite{Guena2012a,PihanLeBars2018} or dark matter \cite{Hees2016,Kobayashi2022,Filzinger2023}. Moreover, overcoming phase noise limitations through advanced microwave signal generation techniques enables more accurate investigations of systematic effects that may lead to improved systematic uncertainties.

For ultrastable microwave generation for the PTB caesium fountain clocks the frequency stability of a cavity stabilized laser is transferred to the microwave spectral range via a frequency comb. For several years, in a former setup an optically stabilized microwave oscillator (OSMW) at 9.6\,GHz was in continuous operation \cite{Lipphardt2017,Weyers2018}, using the frequency comb as a transfer oscillator \cite{Telle2002}. As a replacement of this system, here we describe a new more robust setup for providing an optically generated microwave signal (OGMW), where the 9.6\,GHz microwave signal is obtained directly from a commercial frequency comb, which is locked to a cavity-stabilized fiber laser via high bandwidth actuators \cite{Zhang2012}. In the following Section~\ref{sec_setup} we describe our setup in detail and then present our results in Section~\ref{sec_results}.

\section{Setup}
\label{sec_setup}

As in the previous setup, the source for the short-term frequency stability of the microwave signal is a 1.5\,$\mu$m fiber laser (Koheras BASIK E15) locked by means of the Pound-Drever-Hall technique to an optical cavity made of ultralow expansion (ULE) glass with highly reflecting ULE mirrors \cite{Lipphardt2017}. The resulting laser frequency stability is $\sim$10$^{-15}$ for averaging times in the range of 1 to 10\,s. The output of this cavity-stabilized laser (CSL) is now split and reaches the former and the new frequency comb system via path length stabilized optical fibers.

As a central part of our new setup, we utilise a commercial frequency comb system from Menlo (FC1500-250-ULN). Its femtosecond laser (FSL), with a repetition rate $f_\mathrm{rep}$ of 240\,MHz and a pulse length of 100\,fs, generates an optical comb at a wavelength of 1.5\,$\mu$m with an equidistant mode distribution of about 30\,nm width. Two fast actuators with a control bandwidth of about 1\,MHz and appropriate control electronics allow the stability of the CSL to be transferred to all modes of the comb. The actuators adjust the length and dispersion of the FSL cavity, respectively, to control independently $f_\mathrm{rep}$ and the carrier envelope offset frequency $f_\mathrm{ceo}$.

\begin{figure}[t!]
\centering
  \includegraphics[width=\columnwidth]{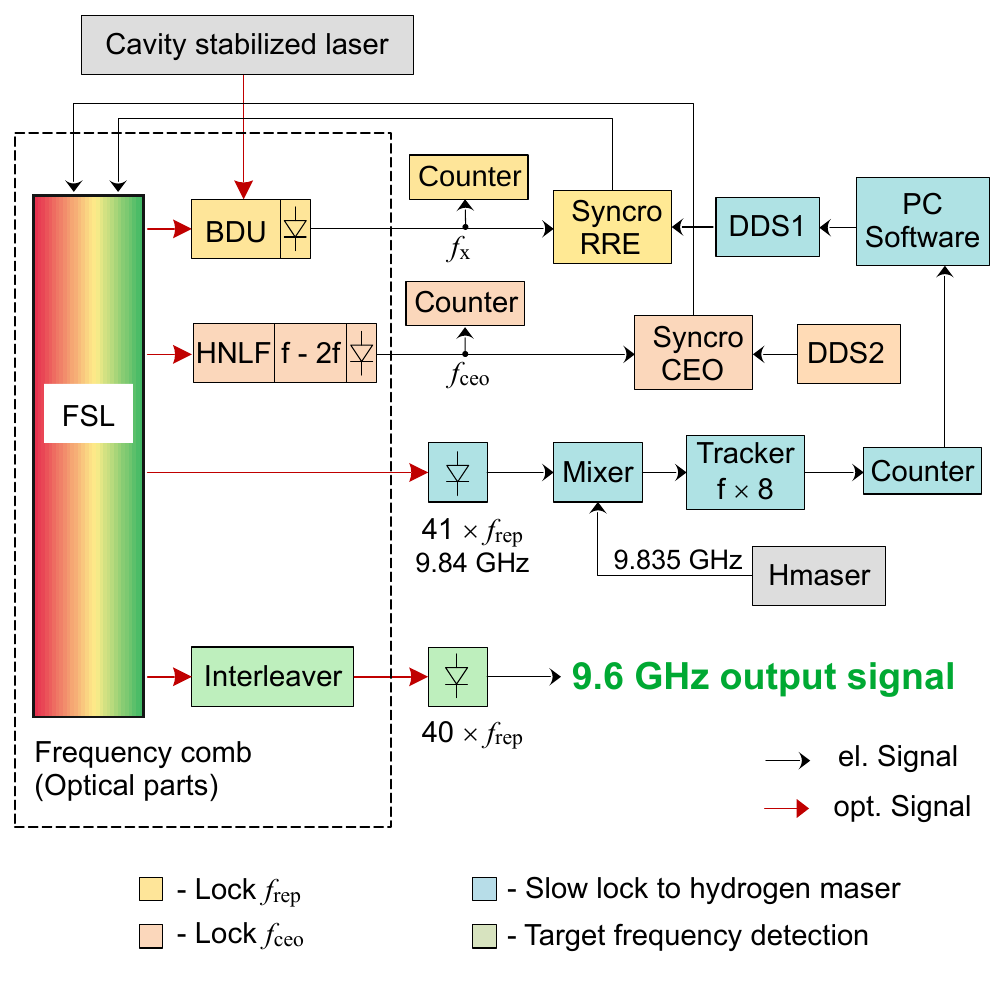}
	\caption{Setup for the optical generation of the 9.6\,GHz microwave signal. The commercial frequency comb system comprises the femtosecond laser (FSL) and its optical accessories (box outlined with dashed line) and two phase discriminators/PI controllers (Syncro RRE and Syncro CEO). CSL: cavity-stabilized laser, BDU: beat detection unit, HNLF: highly nonlinear fibre, DDS: direct digital synthesizer, $f_\mathrm{rep}$: repetition rate, $f_\mathrm{ceo}$: carrier envelope offset frequency, $f_\mathrm{x}$: beat frequency.}
	\label{fig1_Scheme} 
\end{figure}

In Fig.~\ref{fig1_Scheme} both control loops are depicted (beige-colored and peach-colored blocks). To lock $f_\mathrm{rep}$ to the CSL frequency, the comb spectrum is optically pre-filtered ($\pm 0.11$\,nm) with a fiber Bragg grating in a beat detection unit (BDU) where a beat frequency $f_\mathrm{x}$ between one mode frequency of the comb and the frequency of the CSL is generated on an InGaAs photodiode ($\text{SNR}=40$\,dB at 1\,MHz resolution bandwidth). This beat frequency of about 60\,MHz is constantly locked to the frequency of a direct digital synthesizer (DDS1) using a phase discriminator and PI controller. 

For the stabilization of the second degree of freedom of the comb, the offset frequency $f_\mathrm{ceo}$ is extracted by using a highly nonlinear fiber (HNLF), an f-2f interferometer \cite{Jones2000, Holzwarth2000} and an InGaAs photodiode ($\text{SNR}=37$\,dB at 1\,MHz resolution bandwidth). Using again a phase discriminator and PI controller (Syncro CEO), $f_\mathrm{ceo}$ is locked to the 40\,MHz DDS2-frequency, referenced to the output frequency of a hydrogen maser. 

In this "synthesizer mode" of the comb, all optical modes and thus its repetition rate are in fixed frequency relation to the CSL frequency and ideally have the same relative stability. 

The femtosecond pulses of frequency combs generate a comb of microwave frequencies at multiples of the repetition rate on an optical detector. Thereby, the harmonics theoretically extend up to the Fourier limit of about 4\,THz (given by a pulse length of $\sim$100\,fs). In our setup, before detection the electrical signal first passes through an interleaver \cite{Haboucha2011} consisting of a fiber network that duplicates the repetition rate of 240\,MHz four times (green blocks). This technique amplifies the modes that are at a multiple of the frequency 1.92\,GHz and attenuates the shot noise contribution of the others. After detection with a fast InGaAs photodiode (DSC~40S, bandwidth 12\,GHz), the 5$^{\mathrm{th}}$ harmonic (9.6\,GHz) is filtered out with a bandpass filter (9.4-9.8\,GHz) and amplified from about -9\,dBm to 13\,dBm by phase noise specified amplifiers (HMC-C050 and CMD245, not shown in Fig.~\ref{fig1_Scheme}). From the 9.6\,GHz output signal a subsequent homemade synthesis generates the interrogation signal for the fountain clocks \cite{Kazda2020}.

To prevent the comb frequencies and thus the generated 9.6\,GHz signal following the drift of the optical cavity, the repetition rate $f_\mathrm{rep}$ with its harmonics is locked to a hydrogen maser with a time constant of $\sim$50\,s. To realize this time constant, the difference frequency of a harmonic of the repetition rate and a reference frequency from the hydrogen maser is measured by a counter and integrated by a PC (light-blue blocks). The frequency comparison is performed in the microwave range to increase the counter resolution. For this purpose, the 41$^{\mathrm{st}}$ harmonic (9.84\,GHz) is detected with another fast InGaAs photodiode, filtered and mixed down to 5\,MHz first with a reference frequency of 9.6\,GHz and then with a synthesized frequency of 235\,MHz. The 8$^{\mathrm{th}}$ harmonic of the 5\,MHz signal (40\,MHz, virtually @ 78.72\,GHz) is then counted and constantly controlled. For this purpose the drift rate of DDS1 is correspondingly readjusted. The necessary low-noise reference frequencies are obtained from the 100\,MHz output signal of the hydrogen maser. First this signal is filtered by a low-noise 5\,MHz BVA quartz oscillator, from which the 9.6\,GHz reference signal is generated by multiplication. Next the 235\,MHz signal is synthesized using a divider chain, which is similarly employed in our frequency synthesis for the fountain clock interrogation signal \cite{Kazda2020}.

All counters, those for locking $f_\mathrm{rep}$ to the hydrogen maser and those for monitoring $f_\mathrm{x}$ and $f_\mathrm{ceo}$, are synchronous multichannel counters (K+K~FXE \cite{Kramer2004}), which count dead-time free with a resolution of 12\,ps and which are operated in lambda-counting mode with 1\,ms gate time. 

Similar to the former OSMW system, the new OGMW system also offers the possibility to measure the optical clock transition frequencies of PTB's quadrupole ({\em E}\/2) and octupole ({\em E}\/3) $^{171}$Yb$^+$ frequency standards \cite{Tamm2014,Huntemann2016} with respect to each other or the hydrogen maser frequency simultaneously.

\section{Results}
\label{sec_results}

\begin{figure}[t]
\centering
  \includegraphics[width=\columnwidth]{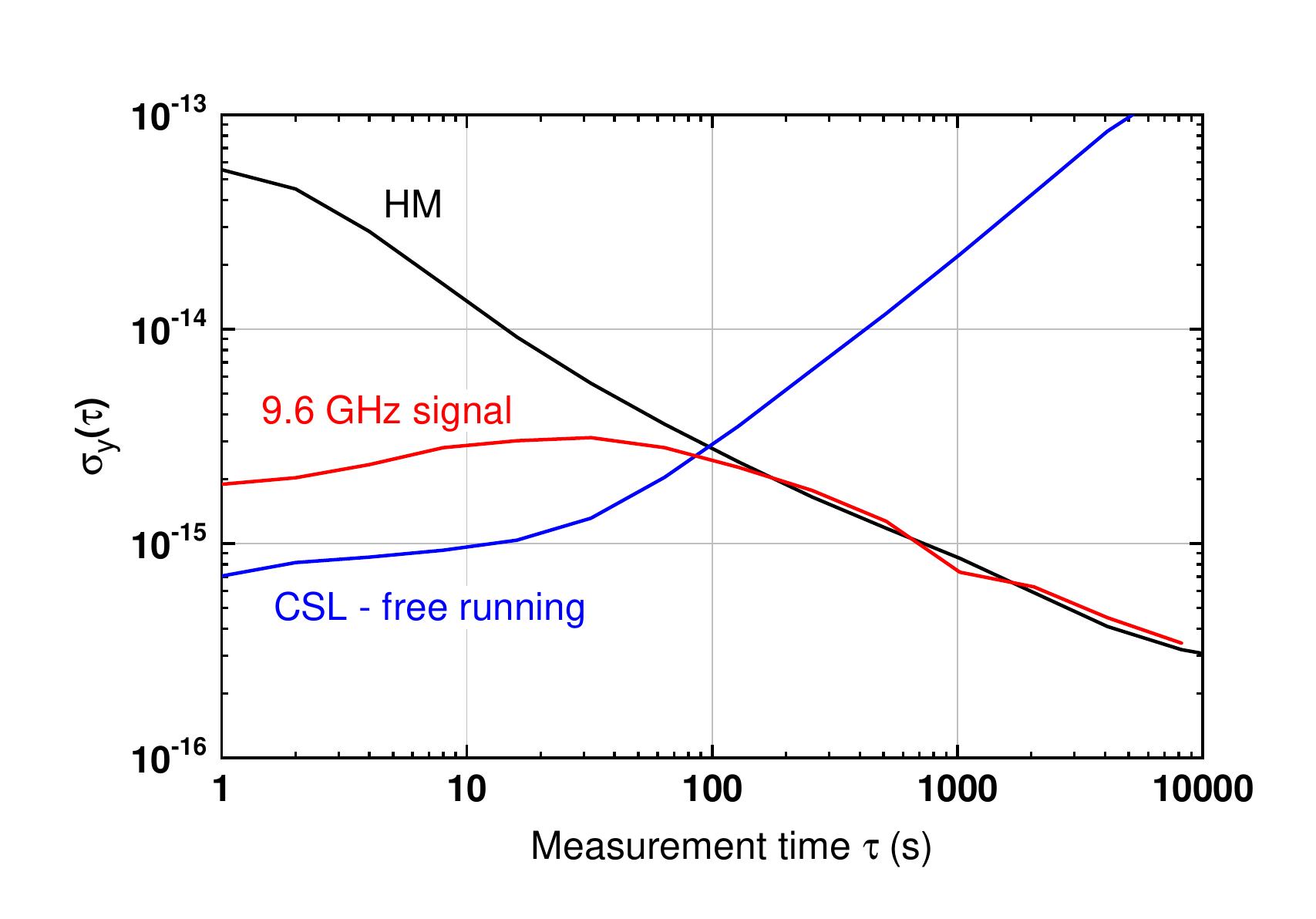}
	\caption{Allan standard deviation $\sigma_y (\tau)$ of the frequencies of the generated 9.6\,GHz signal, the hydrogen maser (HM) and the free-running cavity-stabilized laser (CSL).}
	\label{fig2_AllanCSL} 
\end{figure}

First, in Fig.~\ref{fig2_AllanCSL} the effect of locking $f_\mathrm{rep}$, and thus the generated 9.6\,GHz signal, to the hydrogen maser is visualized by depicting the Allan standard deviations of the microwave signal and the frequencies of the hydrogen maser and the free-running CSL. The data is obtained with another frequency comb by measurements of the ratios of the microwave, the hydrogen maser and the CSL frequencies to the output frequency of an $^{171}$Yb$^+$ single-ion frequency standard \cite{Huntemann2016}, which makes use of a cryogenic silicon cavity \cite{Matei2017} and does not contribute to the measured instability for all Fourier frequencies. The control system is designed to maintain short-term stability on the one hand and to achieve long-term stability given by the maser frequency without control overshoot on the other.

\begin{figure}[t]
\centering
  \includegraphics[width=\columnwidth]{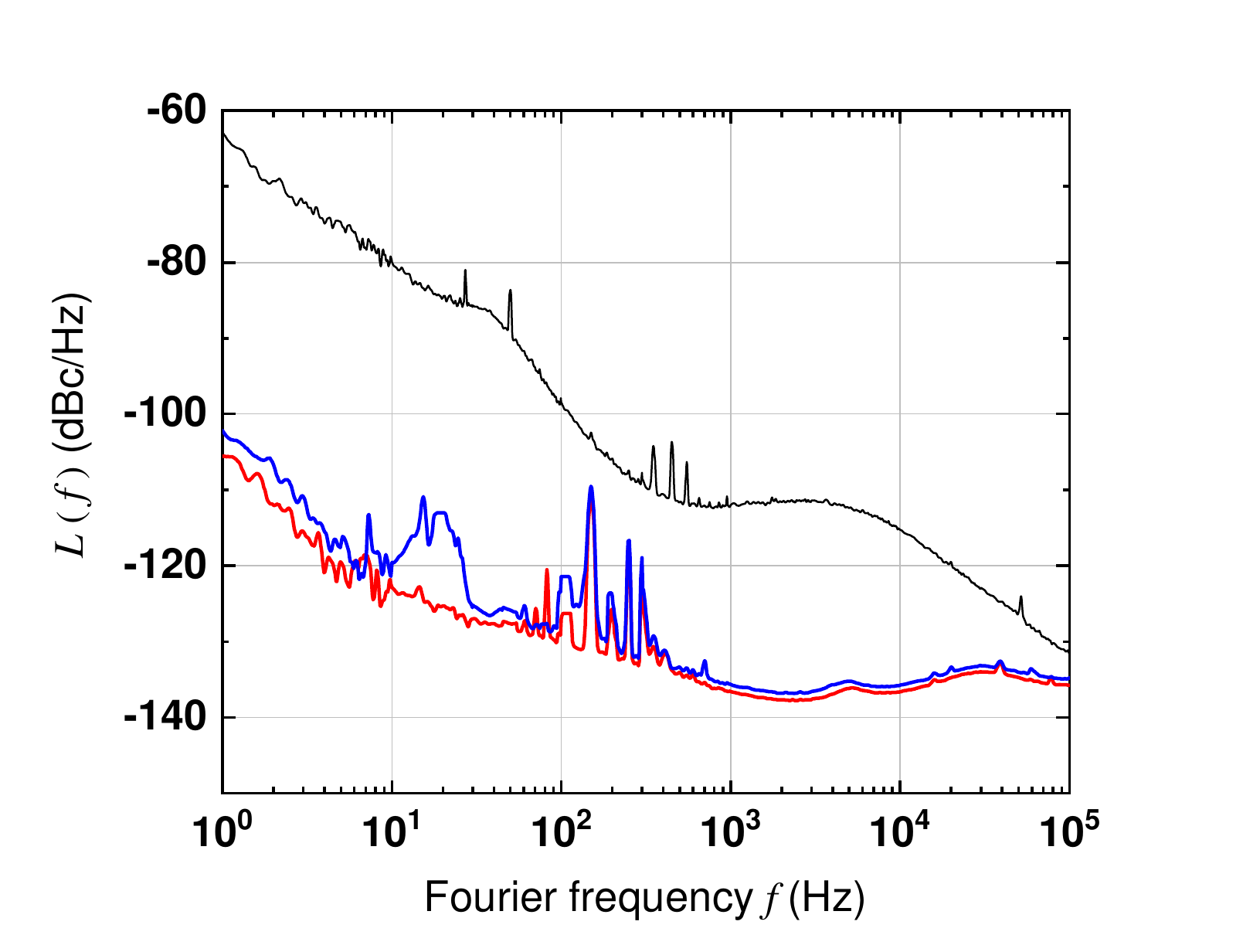}
	\caption{Single-sideband phase noise power spectral density $L(f)$ from comparing 9.6\,GHz microwave signals from the new OGMW and the former OSMW setup. Red: summed phase noise of OGMW and OSMW signals using the same optical reference signal (CSL); blue: summed phase noise of OGMW and OSMW signals using the CSL and a remote cavity stabilized laser (see text), respectively; black: summed phase noise of the signal from a quartz based 9.6\,GHz synthesis and the OGMW signal (for comparison).}
	\label{fig3_PhaseNoise} 
\end{figure}

To characterize the noise performance of the new OGMW setup, two different measurements were performed with respect to the previous OSMW setup. First, a single-sideband phase noise power spectral density measurement was performed to characterize the short-term noise level. Since a direct phase noise measurement close to the carrier is not possible at the present phase noise level, we measured the summed phase noise of the two 9.6\,GHz microwave signals from the OGMW and OSMW setups by subtracting both signals from each other with a mixer, setting a phase difference of the signals of $\pi/2$ for phase-sensitive detection using a phase shifter. The resulting signal is amplified with a homemade low-noise amplifier (42\,dB, 0.7\,nV/$\sqrt{\mathrm{Hz}}$) and 
its phase noise is measured by a phase analyzer (Rohde \& Schwarz FSWP26) in the high-resolution baseband (red line in Fig.~\ref{fig3_PhaseNoise}).

Since the stability of both microwave signals is provided by the same optical reference signal (CSL), this measurement result demonstrates the quality of the frequency stability transfer from the CSL to the microwave signals, using the different techniques of the OGMW and OSMW setups. In our systems, the phase noise of the RF amplifiers and the detection processes of the photodiodes are limiting factors.

To measure the overall phase noise of the microwave signals from the OGMW and OSMW setups, two independent optical reference signals are needed. Therefore, via its frequency comb the OSMW signal was locked to another remote laser, stabilized to a cryogenic silicon cavity \cite{Matei2017}. The frequency stability of the latter laser is about one order of magnitude better than the stability of the CSL. The summed phase noise of both independent microwave signals is depicted as blue line in Fig.~\ref{fig3_PhaseNoise}. The single-sideband phase noise power spectral density $L(f)$ reaches -103\,dBc/Hz at 1\,Hz from the carrier. The additional fluctuations around 10\,Hz are caused by an acoustic sensitivity of the setup for transferring the remote laser light stabilized to the cryogenic silicon cavity between different buildings and do therefore not impact the OGMW signal. For comparison and to demonstrate the benefits of the optical microwave signal generation, in Fig.~\ref{fig3_PhaseNoise} the summed phase noise of a quartz based 9.6\,GHz synthesis and the OGMW signal is shown as well. 

By taking into account the measured single-sideband phase noise power spectral density depicted by the blue line in Fig.~\ref{fig3_PhaseNoise} as an upper limit for the phase noise level of the 9.6\,GHz OGMW signal, we extract the spectral density of the relative frequency fluctuations as $S_y^f(f) = 2.2\times 10^{-30}/f + 8.6\times 10^{-31}/\mathrm{Hz} + 1.7\times 10^{-31}f/\mathrm{Hz}^2 + 8.6\times 10^{-34}f^2/\mathrm{Hz}^3$. From this the upper limit for the frequency instability contribution caused by the Dick effect \cite{Santarelli1998} is calculated for the two caesium fountain clocks CSF1 and CSF2 \cite{Weyers2018}. For the typical operation parameters of the fountains this calculation yields Allan standard deviation contributions $\sigma_\mathrm{y, Dick} (\tau)=1.2 \times 10^{-15} (\tau/1\,\mathrm{s})^{-1/2}$ for CSF1 and $\sigma_\mathrm{y, Dick} (\tau)=1.4 \times 10^{-15} (\tau/1\,\mathrm{s})^{-1/2}$ for CSF2. These noise contributions are negligible for both fountains for normal quantum projection noise limited operation in the $\sigma_\mathrm{y} (\tau)\ge 10^{-14} (\tau/1\,\mathrm{s})^{-1/2}$ Allan standard deviation range \cite{Lipphardt2017,Weyers2018}.

\begin{figure}[b!]
\centering
  \includegraphics[width=\columnwidth]{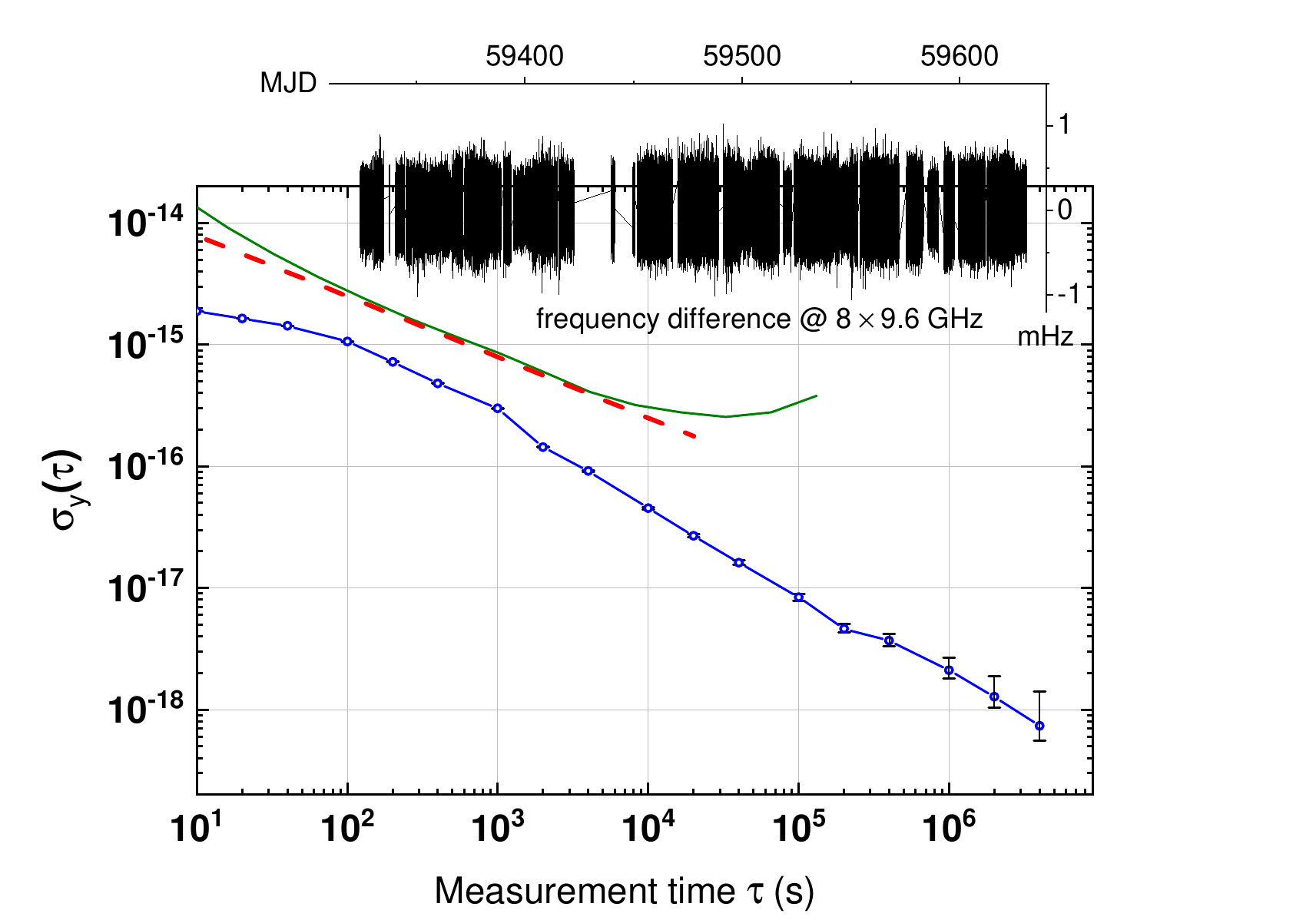}
	\caption{Allan standard deviation $\sigma_y (\tau)$ from measured frequency difference data from the new OGMW and the former OSMW setup (blue data). Indicated in green and red are also the Allan standard deviations of the frequencies of the hydrogen maser and the caesium fountain CSF2 at high atomic density operation. Inset: frequency difference data from the OGMW and the OSMW setup, averaged for 10\,s during a 10 month period, and used for the calculation of the blue data.}
	\label{fig4_AllanMW} 
\end{figure}

In another characterizing measurement, frequency data from the previously utilized OSMW setup, incorporating a 9.6\,GHz dielectric resonator oscillator (DRO), and the new OGMW setup were acquired simultaneously for a measurement period of 10 months. The two associated frequency combs were locked to the same CSL and hydrogen maser. Part of the OSMW setup is an in-loop signal consisting of the frequency difference between the DRO frequency and the 9.595\,GHz comb mode frequency ($f_\mathrm{rep}=252.5$\,MHz) \cite{Lipphardt2017}. For the OSMW and OGMW comparison, this in-loop signal and also the frequency difference between the 9.6\,GHz signal of the OGMW and the same 9.595\,GHz comb mode frequency are registered. From these two intermediate frequency signals the eighth harmonic is generated, again to measure their frequency difference (as 10\,s averages) at high resolution. The resulting data track is shown in the inset of Fig.~\ref{fig4_AllanMW}. The gaps in the recording ($\sim$19\%) are mainly due to failures of the OSMW signal. The data show agreement of the two microwave signals as the measured frequency difference is $0.5 \times 10^{-18}$ with a standard error of the mean of $1.6 \times 10^{-18}$.  

For the same data the Allan standard deviation is plotted for long-term frequency stability analysis (blue graph in Fig.~\ref{fig4_AllanMW}). It has been checked that the Allan deviation at 10\,s measurement time corresponds to the result of the single-sideband phase noise power spectral density measurement (Fig.~\ref{fig3_PhaseNoise}). For comparison, in Fig.~\ref{fig4_AllanMW} the Allan standard deviations of the hydrogen maser and the caesium fountain CSF2 at high atomic density operation \cite{Weyers2018} are shown in green and red, respectively.

\section{Conclusion}
\label{sec_conclusion}

As an alternative for cryogenic oscillators, systems of cavity stabilized lasers and frequency combs (for the optical stabilization of microwave oscillators or the direct generation of microwave signals) have proven to be reliable tools for providing ultrastable microwave signals for the benefit of the frequency stability of caesium fountain clocks. After years of operating an optically stabilized oscillator to generate the microwave signal required by PTB's cesium fountain clocks, this signal is now obtained directly from the femtosecond pulses of a frequency comb. The new setup for the optical generation of an ultrastable 9.6\,GHz microwave signal provides even more robust continuous operation with a phase noise level that is fully compatible with the requirements of fountain clocks.

\begin{backmatter}

\bmsection{Acknowledgments}We would like to thank Melina Filzinger and Nils Huntemann for providing the reference signal from the $^{171}$Yb$^+$ single-ion frequency standard.

\end{backmatter}

\bibliography{KammMW}

\end{document}